# A Robust Frame-based Nonlinear Prediction System for Automatic Speech Coding

Mahmood Yousefi Azar[1], Farbod Razzazi[2]
[1]Department of Computing, Faculty of Science, Macquarie University, NSW 2109 Australia
[2]Department of Electrical Engineering, Science and Research Branch, Islamic Azad University, Tehran
mahmood.yousefiazar@students.mq.edu.au, razzazi@srbiau.ac.ir

**Abstract:** In this paper, we propose a neural-based coding scheme in which an artificial neural network is exploited to automatically compress and decompress speech signals by a trainable approach. Having a two-stage training phase, the system can be fully specified to each speech frame and have robust performance across different speakers and wide range of spoken utterances. Indeed, Frame-based nonlinear predictive coding (FNPC) would code a frame in the procedure of training to predict the frame samples. The motivating objective is to analyze the system behavior in regenerating not only the envelope of spectra, but also the spectra phase. This scheme has been evaluated in time and discrete cosine transform (DCT) domains and the output of predicted phonemes show the potentiality of the FNPC to reconstruct complicated signals. The experiments were conducted on three voiced plosive phonemes, /b/d/g/ in time and DCT domains versus the number of neurons in the hidden layer. Experiments approve the FNPC capability as an automatic coding system by which /b/d/g/ phonemes have been reproduced with a good accuracy. Evaluations revealed that the performance of FNPC system, trained to predict DCT coefficients is more desirable, particularly for frames with the wider distribution of energy, compared to time samples.

## 1. Introduction:

Speech signals have unique attributions compared to other signals (e.g. audio, music signals) and this uniqueness demands efficient techniques. Recently, new applications, (e.g. cellular telephony, mobile satellite communications, audio for video teleconferencing systems) have ignited scholars' interests to investigate more efficient approaches specifically designed for speech signals and online purposes.

One subset of the most popular techniques among speech coding models is linear predictive coding (LPC) and its variants. The idea of LPC modeling, inspired from human speech production system, is that the speech signal can be predicted using a linear combination of its past samples. Although scientists have been conducting many studies to modify LPC's weak points, its linearity pre-assumption have restricted the performance. LPC is, in fact, based on a exceedingly simplified version of the human speech apparatus. For example, it assumes the source signal and the filter are independent phenomena from each other. Additionally, an autoregressive all-pole filter has been replaced with human vocal track, neglecting the zeros of the filter [1]. This study is on the basis of the non-linear predictive coding as a generalized model of LPC as a non-linear vocal track representation [2]. LPC is not alone in coding approaches and other techniques, based on the human auditory system and statistical modeling, play a significant role in coding aims. For example, Mel-frequency cepstral coefficient (MFCC) feature vectors can be used to model the power spectral density envelope, thereby being useful for compression speech data. However, the algorithm of MFCC is based on the power spectrum of signal, thus losing the phase of the signal is inevitable, and it may affect human perception. In fact, there is concrete evidence of phase information usefulness in human perception [3-5].

Using artificial neural networks for speech feature extraction dates back to the 80's when Tishby et al had work on the speech signals [6]. The intuition about the nonlinearity of human speech apparatus in elaborately producing speech signals suggested that a nonlinear and dynamic structure can be more appropriate compared to linear and static one. In 1988, Waibel proposed a new topology, a time delayed neural network, which was more adaptable to handle dynamic and correlated signals [7]. The core of this technique was a multilayer perceptron (MLP) network to which temporal sequences were delayed before getting applied. In recent years, promising studies have being done to use neural networks for speech representation. This approach has been extended to other artificial intelligent tools for feature selection [8-11].

In addition to the mentioned neural-network-based systems, there have been other promising methods such as deep neural networks, convolutional neural networks, etc. One of the main features of these methods is to use deep learning algorithms. Having more than one hidden layer and independent training phases, they are able to effectively employ for speech processing purposes [12-14].

In this paper, we are going to present an effective topology to code the speech signals. The Frame-based neural predictive coding (FNPC) is primarily designed to be adapted into each frame of speech signal and to predict, automatically, a sample from a nonlinear combination of its finite past samples. The FNPC system should be able to learn a wide variety of acoustic features during first training phase. Having trained adequate parameters, the FNPC will be able to specifically learn a frame of speech, either the original time domain signal or its discrete cosine transform (DCT) coefficients. Since the structure has only one hidden layer, it cannot be considered as a deep neural network; however, its independent training procedures can be analogous with those networks. The proposed method objective is to ease the complexity of classic NPC and present a robust system to reproduce the original signal, in addition to an estimation of a frame's spectral envelope. The technique can be planned based on the number of created codes versus the quality of the reconstructed signal.

The structure of the paper is as follows. There will be a section to provide a briefly overview of the topology and relations in FNPC. The third section belongs to preprocessing formulations where the relations for transforming time series into DCT domain and to make normalized inputs have been presented. In section 4, after describing the experiments setups, the performance of FNPC has evaluated using time and DCT domain samples versus increasing the number of producing codes, separately. Finally, section 5 is to draw a conclusion.

## 2. Neural-based Predictive Coding

Among the techniques which have been inspired from human vocal track system, LPC has been presented as a fundamental speech feature extraction structure in the purpose of speech synthesis, recognition and coding. The gist of the system is that if short-time windowed speech signals can be assumed as a short-time stationary signal, human vocal track may be modeled with an autoregressive all-pole filter. The filter parameters would be used as the extracted features of the speech frame and they can be utilized for reproducing the signal, recognition purposes, etc. LPC, by which the speech production apparatus is highly simplified, can only estimate the spectral envelope of the frame approximately. In other words, a speech frame spectrum is primarily characterized only by formant peaks and it leaves out the valleys of the spectrum.

Lapedes was one of the first researchers to put forward a neural network as a nonlinear predictor [15]. From this perspective, a one layer feed forward neural network with L inputs and one output, where L stands for the prediction window, is indeed a linear predictor. To adapt the LPC idea into a non-linear predictive coding, a multi-layer non-linear feed forward neural network to predict the each speech sample from L previous samples would be substituted for the LPC filter.

### 2.1. FNPC Topology

The Frame-based neural predictive coding, consisting of a multi-layer perceptron is a trainable coding method. The codes are the weights of the output layer. In FNPC, there are two training phases for an MLP with two weight layers. First phase, as the mapping phase, is dedicated to a nonlinear mapping of data and all speech samples would be used for training the first layer weights of an FNPC network. The second stage, named as coding phase, is the encoding procedure when the network tries to learn each frame separately, while the first layer weights are assumed to be constant. Indeed, FNPC's trainability comes from the second phase. The phases are shown in figure 1 and 2.

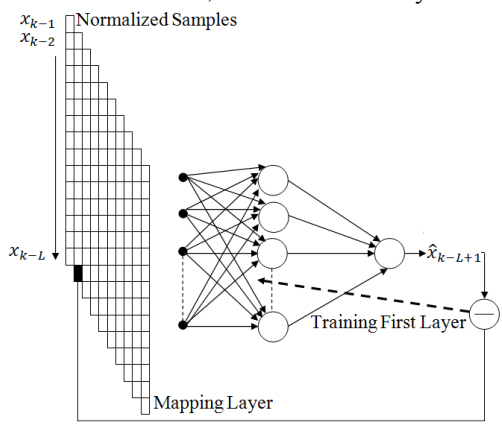

Fig. 1 The mapping phase of FNPC to train first layer weights

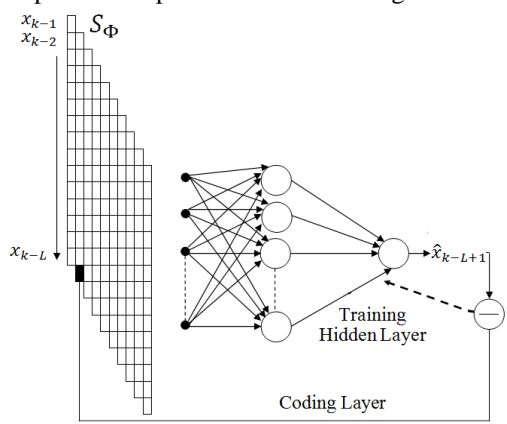

Fig.2 The coding stage of FNP where the network is adopted into the $\Phi^{th}$ frame

To be more exact, in the mapping phase, a rough predictor is trying to accommodate itself to the whole speech properties and make a non-linear mapping to reduce the dimension of input vectors. Evidently, this adaptation will not be special for every speech frame, but a general speech adjustment. In the second training phase, the predictor is matched with a specific frame supposing every sample of the frame might be reproduced by exerting these nonlinearly produced codes in the decoder network. In this trend, the target of optimization is to minimize MSE (mean square error) of the output while the weights of the first layer of the networks are fixed. The optimized weights are regarded as the frame's code, because they can put into the decoder to reconstruct the frame.

When it comes to compare the speed of two training phases, it is true that the second phase is always faster than the first phase, since the number of neurons in the hidden layer, the number of codes, is less than the number of the first layer weights but they should be trained for each frame; thus, the most of complexity of computation is due to the second phase. This computation complexity depends on the number of neurons in the hidden layer which affects the intelligibility of the reconstructed signal.

## 2.2. FNPC Formulations

This section, by which the mathematical formulations of FNPC training phases are presented, is similar for both time and DCT domains [16, 17].

Let's assume a substitution of a non-linear predictive function, a three layer MLP with one output neuron called the prediction neuron, for the linear one. The general function would be $y_{k+1} = \phi^2\left((w^2)^T\left(\phi^1(w^1\hat{X}_k)\right)\right)$ where $\hat{X}_k$ denotes the standardized vector with respect to mean and standard deviation of the input samples. The $w^1 = [\ldots, w_i^1, \ldots]^T$ is the matrix of the first layer weights, $w^2 = [\ldots, w_i^2, \ldots]^T$ is the vector of the prediction neuron cell weights, $\phi^1$ and $\phi^2$ are the first and second layers activation functions respectively. In FNPC model, the $w^1$ weights are the mapping parameters, estimated at the first phase to make a nonlinear mapping for all speech signals' attributes. This stage results in a reduction in the data dimension. This stage will be remained constant in the second stage. Figure 1 shows the non-linear mapping phase, dedicated to the whole database samples. The estimation procedure is performed by using a back-propagation (BP) procedure to train both layers. The training sequences of the samples are presented to the network without any framing preprocessing in time domain. However, normalization which is applied for the whole signal samples would take place in this phase.

The second stage procedure is in fact a routine which would be used in online speech processing. The coding phase has illustrated in figure 2. The second weight layer (i.e. codes), is adapted by the frame samples using a gradient descent based algorithm what is a commonly used algorithm for neural network training. Assuming M as the output vector size (i.e. the number of codes), we will have M-1 neurons in the hidden layer. Although the bias of the output neuron can be remained constant, it was allowed to be adapted. Any frame with length N, can create an N-L length vector to train, where L is the predictive window length. If we represent $S_\Phi$ as the total number of samples in the $\Phi^{th}$ frame, there are:

$$S_\Phi = \{X_1, X_2, \ldots, X_{N-L}\} \tag{1}$$

$$X_n = [x_{k-1}, x_{k-2}, \ldots, x_{k-L}]^T \tag{2}$$

For every sample of $X_n$, system is trained to predict $y_k$ as the next sample. To train the network in both phases (all the database), the error of BP algorithm should be adapted. The assumed error criterion is mean squared error.

$$\hat{F}(k) = \sum_{K=1}^{N-L}(e_k)^2 \tag{3}$$

where $e_k = y_k - \hat{y}_k$ is the prediction error. As the output of the $\Phi^{th}$ frame is calculated through the relation:

$$\hat{y}_k = \varphi(v_k) = \varphi\left(\sum_{i=1}^{M-1}\varphi^1(w_i^1)Z_{i,k} + b^2\right) \tag{4}$$

where $w_i^1$ are the weights of the first layer, (i.e. mapping matrix), trained for all database. They would be remained constant in the coding phase.

In BP formulation, the output of the first layer is:

$$Z_{i,k} = \varphi(v_{i,k}) = \varphi(\sum_{j=1}^{L} w_{i,j}^1 x_{k-j} + b_i) \tag{5}$$

The BP relations can be expressed as:

$$\Delta w_i^{2\Phi} = 2\lambda \frac{\partial \varphi^1}{\partial w_i}(w_i^1) \sum_{k=1}^{N-L} e_k \frac{\partial \varphi}{\partial v_k}(v_k) Z_{i,k} \tag{6}$$

where i is the number of parameters of $\Phi^{th}$ frame and $\lambda$ is the learning rate.

After enough number of iterations in the mapping phase and finding the weights of the first layer, we should store these weights to use at the coding phase. At the coding stage, the output layer weights should be modified to minimize the following criteria:

$$\hat{F}(k) = \frac{1}{2} \sum_k \left( y_k - \varphi\left( w_i^{2\Phi} x_k \right) \right)^2 \tag{7}$$

The mentioned criteria are a single layer perceptron cost function. Therefore, the weights modification relation would be the same as a perceptron.

## 3. Data Transformation

In addition to the above mentioned formulations, other relations have to be considered for the system with DCT domain input. The following is the transformation function:

$$Y(k) = \begin{cases} \sqrt{\frac{2}{N}} \sum_{n=1}^{N} x(n) \cos \frac{\pi(2n-1)(k-1)}{2N} & k = 2, \ldots, N \\ \frac{1}{\sqrt{N}} & k = 1 \end{cases} \tag{8}$$

Here, N is the window width, $x(n)$ is the sequence of samples of one windowed frame of the speech signal and $Y(k)$ is DCT coefficients of $x(n)$.

One of the most important features of DCT transformation is that it will create a real sequence and totally restorable transform for speech frames. For real and even signals, DCT is equivalent to fast Fourier transform. Therefore, the FNPC system with DCT input would potentially be able to learn all parameters by which the original speech signal can be reproduced. In fact, we are directly going to produce the spectrum of the signal, not only an estimation of the signal's spectral envelope.

Another characteristic issue of this transform is considerable compaction of the signal energy into lower coefficients. It has been shown that classic and warped DCT features are compact well and are assumed as a good representation of the speech samples [18-20]. This may contribute to select a range of coefficients, mostly lower ones, by that means smaller matrix should be learned in the coding stage. Despite that, we have used all transformed values.

As the range of values of raw data varies widely, in many machine learning algorithms, standardizing can improve the effectiveness of the training stage. One of the most common techniques for normalization is the standardization the whole dataset to have zero mean and unit standard deviation [21, 22]. Indeed, this linear transformation will be implemented over all speech frames in the mapping phase. This adaptation is presented as:

$$\acute{x}_{k+1} = (x_{k+1} - \mu_{k+1})/\sigma_{k+1} \qquad 0 \leq k \leq L - 1 \tag{9}$$

where $\acute{x}$ is the normalized sample, x is the original sample and L is the predictive window length. µ and σ are the mean and standard deviation across all database samples respectively, estimated as:

$$\mu_{k+1} = \frac{1}{S-L+1} \sum_{j=k+1}^{S-L+1} x_j \qquad 0 \leq k \leq S - L \tag{10}$$

and

$$\sigma_{k+1} = \frac{1}{S-L+1} \sum_{j=k+1}^{S-L+1} (x_j - \mu_{k+1})^2 \qquad 0 \leq k \leq S - L \tag{11}$$

where S is the total samples. This transformation will be used in the mapping phase for both time and DCT domains whereas this preprocessing is not required in the coding phase.

## 4. Experiments and Results

### 4.1. Experiments Setup

The evaluation benchmark is TIMIT corpus, standardized by the National Institute of Standards and Technology (NIST). The database comprises eight common North American dialects. 630 speakers, both genders, have uttered 10 phonetically transcribed and labeled sentences, where first and second statements are the same for all speakers. The proposed method was evaluated on the first (New England) dialect [23].

It contains 61 phones and 39 phonemes and the whole samples of 61 phones are employed to train the mapping phase. The tests were conducted on voiced plosive phonemes, (i.e. /b/, /d/, and /g/) in the coding phase. The special statistical property of these phonemes and organizing a criterion for subjective testing are the reason to select this subset to evaluate the proposed system. The stressed, syllable-initial phonemes like /b/d/g/ have the transient features and widely used in speech processing evaluation systems [24]. Given the frequency spectrum of these phonemes, the energy of phoneme /b/ is distributed in the wide range whereas most of the energy for phoneme /d/ is in low frequency and the attributes of phoneme /g/ are somewhere between /b/ and /d/.

The first step in preprocessing the speech data was windowing. To have a fully restorable signal, the experiments have carried out using the Hamming window with 256 samples length and 50% overlap. The length of the predicting widow was 40 in both time and DCT domains and the resulted data was normalized to have zero mean and unit variance. This normalization may improve the neural network performance in the mapping phase. One frame samples are collected from the subset in the depicted figures.

### 4.2. Temporal Domain Prediction

To analyze FNPC capability in coding a speech signal, we tested FNPC model by increasing the number of neurons in the hidden layer with time samples as the input signal. It is true that predictive coding techniques are generally trying to minimize the RMS error, but the human auditory system does not perceive using this criterion. figure 3 illustrates the original and predicted /d/ and /g/ in time domain versus the number of neurons. It is evident that escalating the number of neurons in the hidden layer enhances the systems to reconstruct more intelligible voice. It appears that very complicated phonemes with overlapping frequency characteristics distribution at the frequency domain can be reproduced by the proposed topology. Also, there can be a trade-off between the accuracy of predicted signal and the number of codes, thereby reducing the computational complexity and memory requirements.

LPC modeling is on the basis of time domain, having said that, LPC modeling tries to only estimate signal's spectral envelope in the frequency domain. However, FNPC is being designed to learn the original signal in time domain and it can store the signal's phase in addition to power spectral density in its modeling. This happens due to specialized training phase for every frame in the coding phase.

When it comes to reproducing /d/ and /g/ phonemes, FNPC has an excellent output with time samples; however, as it is shown in figure 3, the network with 8 and 16 neurons in the output layer cannot model a complicated phoneme like /b/ as convincing as /d/ and /g/. In truth, FNPC is able to predict signals in which the main features are in the lower frequencies. In other words, it is not able to learn signal that have sharp changes and slight changes, simultaneously, over a short period of time. To depict this issue, the decoded signal of /b/ and its frequency response have been illustrated in figure 4. It has 11 neurons in the hidden layer.

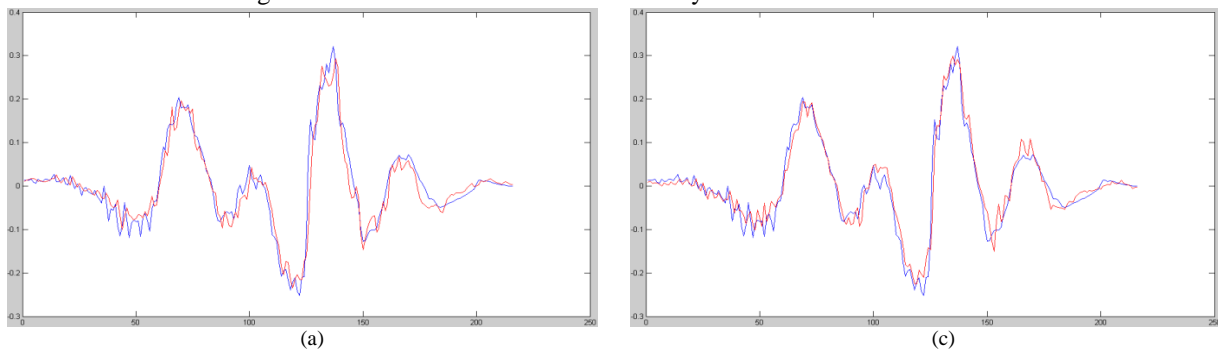

(a)      (c)

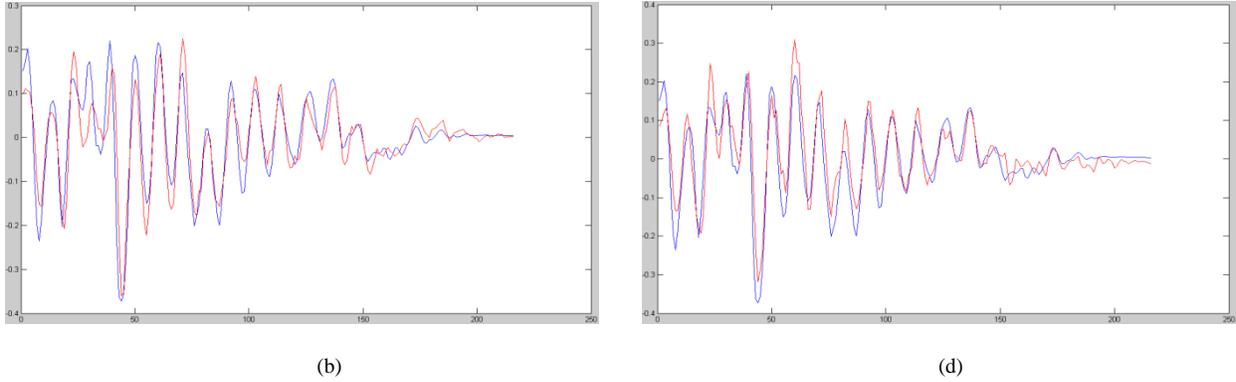

Fig. 3. The FNPC behavior to predict time samples. Blue original signal, red: predicted signal.
a: phoneme /d/ with 8 neurons in coding layer
b:phoneme /g/ with 8 neurons in coding layer
c: phoneme /d/ with 16 neurons in coding layer
d: phoneme /g/ with 16 neurons in coding layer

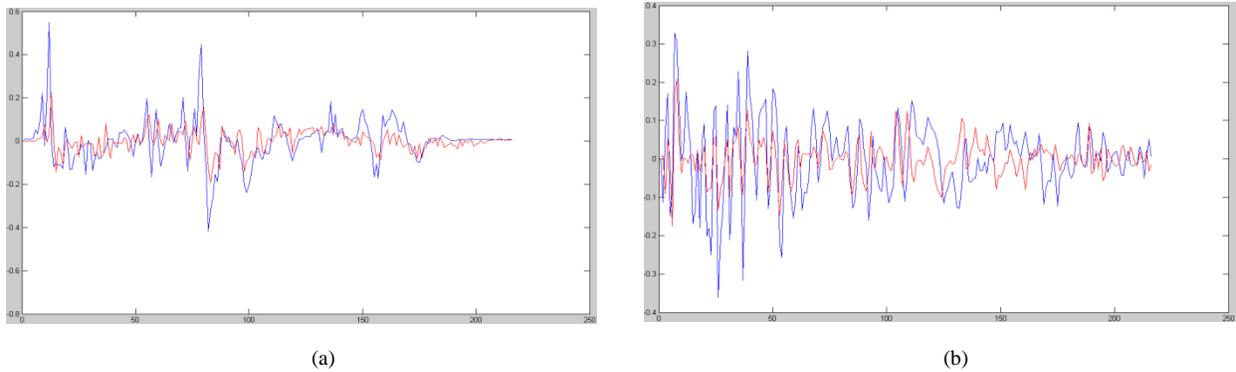

Fig 4. The output of system with 11 neurons in the hidden layer. Blue: original signal, red: the predicted signal of phoneme /b/. (a): time domain and (b): DCT domain

*4.3. Spectral Prediction*

In this experiment, we have arranged a set of simulations to evaluate the potentiality of FNPC in predicting speech signals using DCT samples for prediction. The idea is trying to propose a method to reproduce the original signal spectrum, not only the envelope. That is to say, LPC modeling behavior in frequency domain is based on the formants peaks and it does omit valleys but the proposed technique will exploit the DCT coefficients to reproduce both formant peaks and spectral valleys.

Figure 5 clearly illustrates the FNPC ability to precisely reconstruct DCT coefficients of /d/ and /g/ phonemes versus the number of codes, (i.e. the number of neurons in the hidden layer). Similar to time domain, the FNPC performance in predicting DCT coefficients improves with increasing the coding layer's neurons. Using DCT domain as an input in FNPC is superior to the time domain samples for phoneme /b/ and it may come from the fact that since a signal's high frequency features is equal to high changing rate in time domain, FNPC is not able to predict such changing rate over its predictive window in time domain.

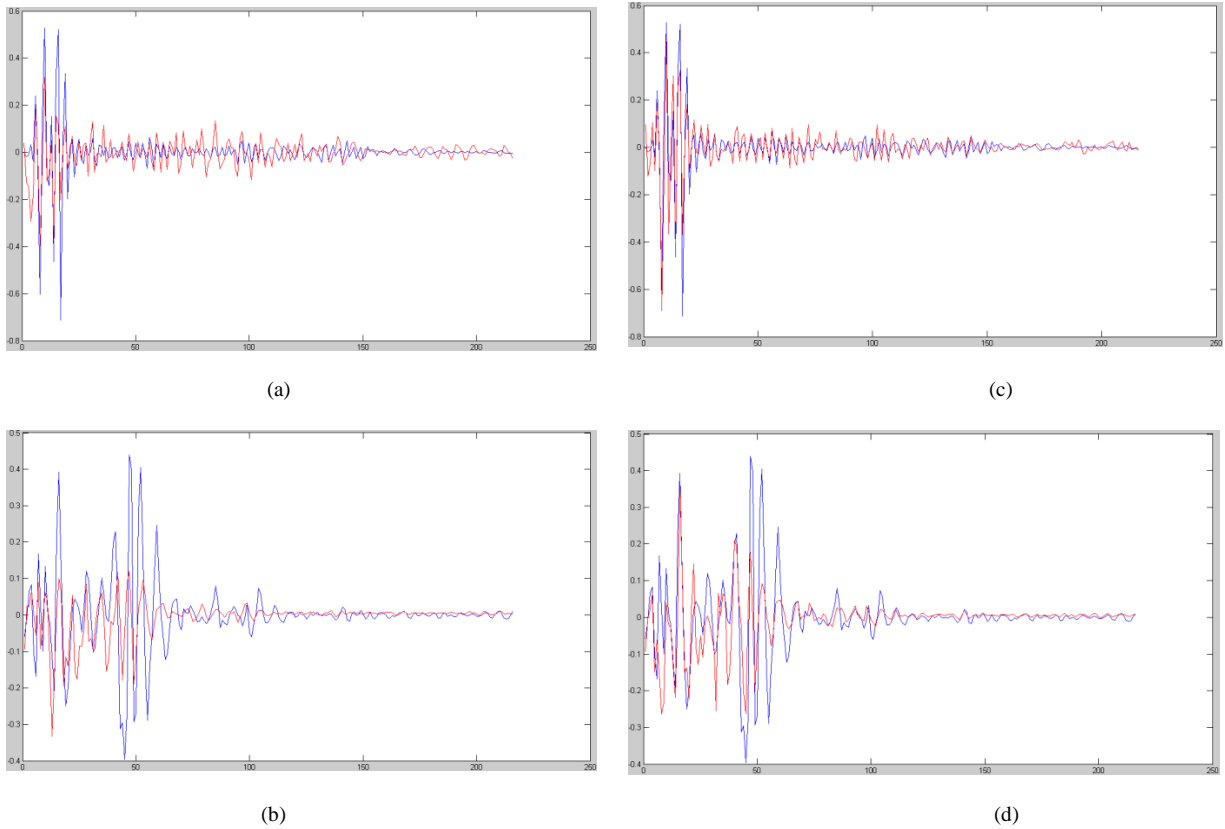

Fig 5. The FNPC behavior to predict DCT samples. Blue: original signal, red: predicted signal.
a: phoneme /d/ with 8 neurons in coding layer
b: phoneme /g/ with 8 neurons in coding layer
c: phoneme /d/ with 16 neurons in coding layer
d: phoneme /g/ with 16 neurons in coding layer

For a complicated signal like /b/ prediction of DCT coefficients using a FNPC with 11 neurons in coding phase is more precise compared to time samples. This issue has presented in figure 6. However, the weak point of FNPC in DCT domain compared to time domain is in frames, in which the density distribution of samples is uneven across the frame.

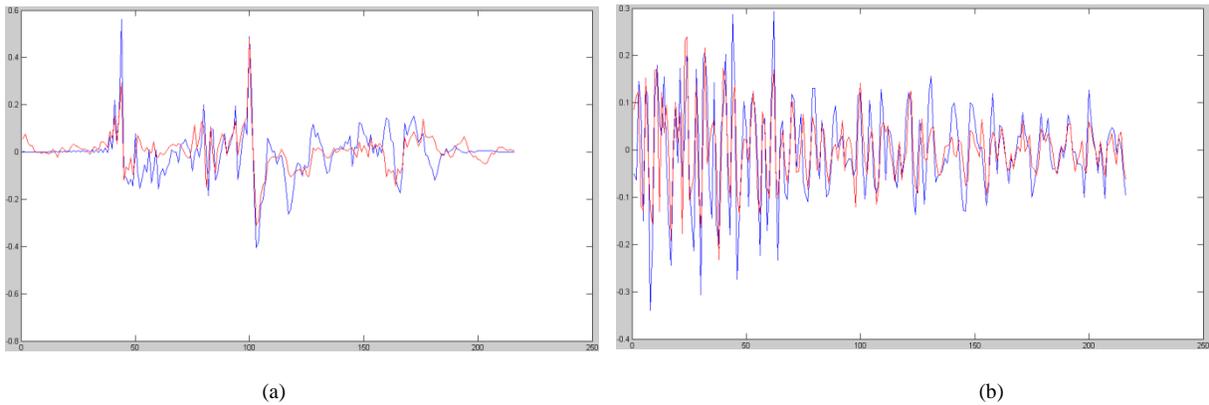

Fig 6. The output of system with 11 neurons in hidden layer. Blue: original signal, red: predicted signal of phoneme /b/. (a) time domain and (b) DCT domain

*4.4. Continuous Speech Intelligibility*

As FNPC is designed to encode continuous speech, we conducted commonly used objective tests in addition to assessing the spectrogram of a continuous speech signal. To evaluate speech intelligibility, four benchmarks were evaluated: Perceptual Evaluation of Speech Quality (PESQ) which is based on human auditory perception and was officially standardized by the International Telecommunication Union, Telecommunication Standardization Sector (ITU-T) as standard P.862 in February 2001, segmental signal-to-noise ratio (SegSNR) as a waveform-based benchmark, log-likelihood ratio (LLR) and weighted spectral slope (WSS), both as spectral-based measures [25][26]. Indeed, PESQ and SNRseg show the quality of predicted signal while LLR and WSS results can be considered as an index when FNPC codes may be used in a speech recognition system. More precisely, LLR and WSS are placed on the table for presenting FNPC's potentiality to produce classification features for future works. The objective of these experiments is to show the performance of FNPC for clean and noisy signals in both time and DCT domains.

Tables 1 and 2 present the results of the tests conducted on a sample TIMIT wave file, TEST\DR1\MDAB0\SI1039.wav, predicted in time and DCT domains [12]. As can be seen, having increased the number of neurons in hidden layer, PESQ results will improve in time domain, comparable even with enhanced signals [27]. Normally, the network can learn signals in more details when the number of weights was increased. However, this trend is not the same for DCT domain and it may come from the distribution of signals in the domain. Most speech signals have an uneven distribution of energy, based on DCT representation. More clearly, when a frame of speech signal in DCT domain is chosen to be the networks' input, in coding phase, the network must learn samples with wide range of amplitude, most of which in lower frequency. This disproportional input may cause more challenge for the network to be adapted for prediction, thereby reducing the quality of reconstructed signal.

Additive white Gaussian noise (AWGN) is commonly used to evaluate a system. The noise was added to the tested speech signal at SNR = 20dB. Table 3 shows the output of FNPC for noisy signal in both time and DCT domains. PESQ is quite impressive for DCT domain and this improvement might be an experimental justification for above mentioned reason. After adding AWGN to the clear signal, the distribution in DCT domain can be smoother and this factor will help the network learn more efficient. On the other hand, the quality of the predicted signal in time domain is impaired due to time sample complexity while adding AWGN.

SNRseg for time domain is generally less than DCT domain for both clean and noisy signals. One explanation may lie in selection of more suitable domain to predict most phonemes of the signal.

Table 1. The results of objective quality measures for time domain

| The Number of neurons | PESQ | SNRseg | WSS | LLR |
|---|---|---|---|---|
| 8 | 3.67 | 9.00 | 10.2 | 0.17 |
| 11 | 3.77 | 9.26 | 9.7 | 0.20 |
| 16 | 3.80 | 9.60 | 9.8 | 0.18 |

Table 2. The results of objective quality measures for DCT domain

| The Number of neurons | PESQ | SNRseg | WSS | LLR |
|---|---|---|---|---|
| 8 | 3.77 | 14.9 | 2.2 | 0.14 |
| 11 | 3.78 | 14.6 | 2.5 | 0.15 |
| 16 | 3.63 | 14.8 | 2.6 | 0.15 |

Table 3. The results of objective quality measures conducted on TEST\DR1\MDAB0\SI1039.wav file with 20dB AWGN

| | PESQ | SNRseg | WSS | LLR |
|---|---|---|---|---|
| **Time domain** | 3.10 | 1.7 | 16.6 | 0.19 |
| **DCT domain** | 4.33 | 9.4 | 1.1 | 0.06 |

In addition to objective tests, the spectrogram of the sample has shown in figure 7. It can be observed that FNPC can predict the original signal with high accuracy both in time and DCT domains. Also, figure 8 demonstrates the effect of signal reconstruction in the presence of noise in the spectrogram of the signal. As it can be seen, although the DCT domain reconstruction is more successful to reconstruct the noisy signal, there are some de-noising clues revealed in time domain reconstruction.

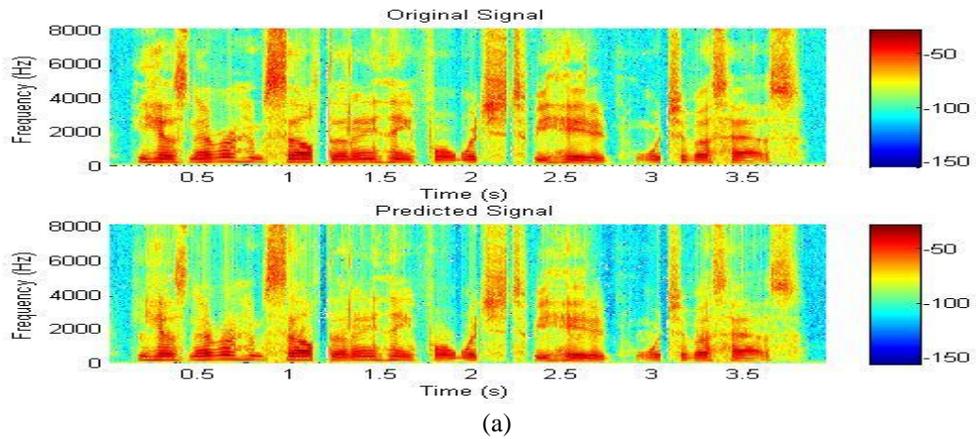

(a)

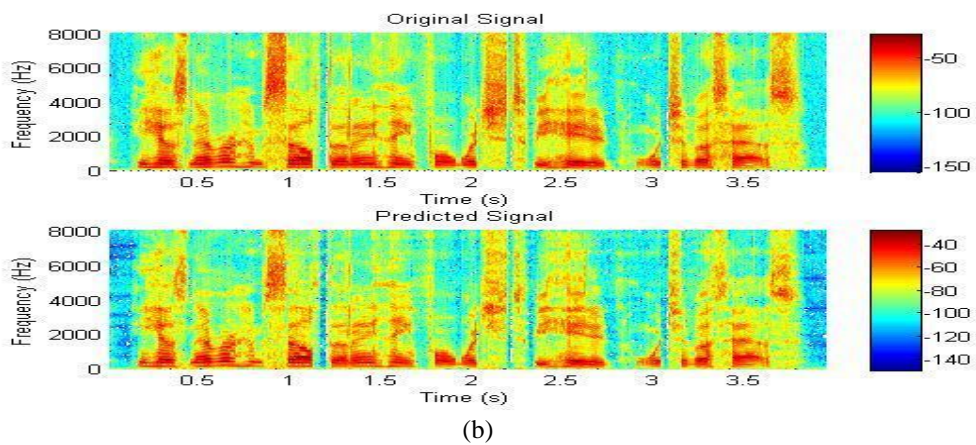

(b)

Fig 7.The spectrogram of the original and predicted TEST\DR1\MDAB0\SI1039.wav file from TIMIT database.
(a) DCT domain and (b) time domain

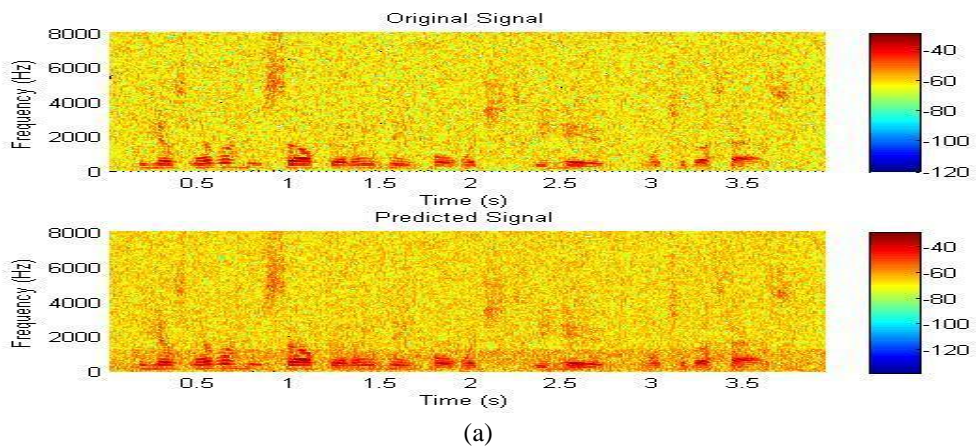

(a)

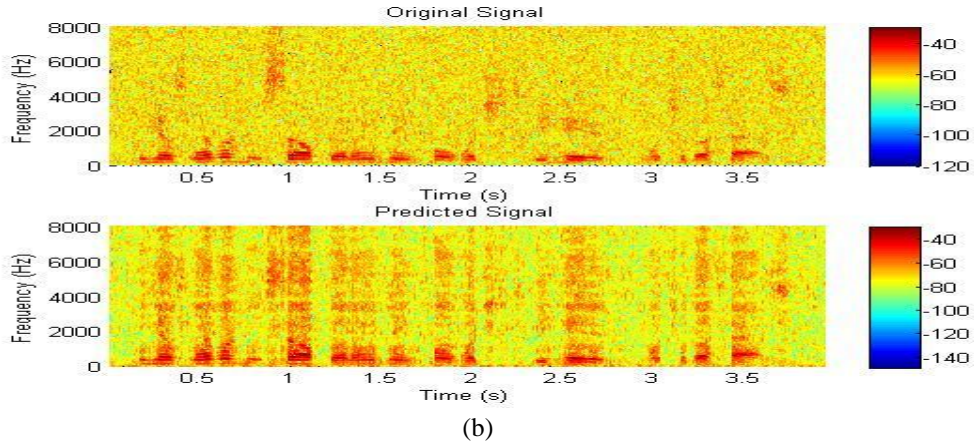

(b)

Fig 8. The spectrogram of the noisy signal, 20 dB, and predicted signal. (a) DCT domain and (b) time domain

## 5. Conclusion

Designing a desirable speech coding system has long been a concern of the speech processing researchers. We proposed a nonlinear predictor machine to automatically code the speech signal. The restored frame of signal would be precisely reproduced. The proposed method is specifically adaptable for each frame using two training phases. Since both encoding and decoding procedures are a fully neural-network-based scheme, a suitable hardware would facilitate this online algorithm.

Experiments have shown the system's capability in reconstructing a frame of speech signal. In fact, FNPC does not pre-assume linear predictive coding restrictions. Therefore, by using the original signal in time domain, FNPC can nonlinearly predict each sample of the signal with a high accuracy. Having the time domain as the input of the system, it can store significant features, appropriate for reconstructing the signal. Although the complexity of some speech phonemes (e.g. the phoneme /b/) can dim hopes of finding a desirable coding scheme, the output quality measured with subjective tests using DCT coefficients shows FNPC efficiency. In addition to signals in which most of the energy is in lower spectra, the proposed method is able to code complicated signal produced by human vocal track. As it is presented, the number of codes for both time and DCT domains is effective on the intelligibility of decoded signals contrary to classic techniques. Experiments show FNPC's ability to code continuous speech using time samples and DCT coefficients and the proposed system in DCT domain is more robust to noise compared to time domain. Working on deep architectures and the influence of predictive windows may enhance the FNPC's efficiency.